\documentstyle[12pt]{article}

\newcommand{\ket}[1]{\left | \, #1 \right \rangle}
\newcommand{\bra}[1]{\left \langle #1 \, \right |}

\newcommand{\beq}{\begin{equation}}
\newcommand{\eeq}{\end{equation}}
\title{Quantum Error Correction for Communication}
\author{Artur Ekert and Chiara Macchiavello\\
Clarendon Laboratory, University of Oxford}
\date{February 1996}

\begin{document}
\maketitle
\begin{abstract}
We show how procedures which can correct phase and amplitude errors
can be directly applied to correct errors due to quantum entanglement.
We specify general criteria for quantum error correction,  introduce
quantum versions of the Hamming and the Gilbert-Varshamov bounds and
comment on the practical implementations of quantum codes.  
\end{abstract}

Suppose we want to transmit a block of $l$ qubits (i.e. two-state
quantum systems) in some unknown quantum state (pure or mixed) over a
noisy quantum channel.  Here `noisy' means that  each transmitted
qubit may, with some small probability $p$, become entangled with the
channel. In order to increase the probability of the error-free 
transmission we can encode the state of $l$ qubits into a set of $n$
qubits and try to disentangle a certain number of qubits from the
channel at the receiving end. This paper specifies  conditions under 
which such encoding and disentanglement are possible.

Let us start with introducing convenient definitions and notation.
Amplitude errors in a block of $n$ qubits are defined as a sequence of
$\sigma_x$ transformations performed
on qubits at locations specified by a binary n-tuple $\alpha$
(non-zero entries of $\alpha$ mark the locations of the affected
qubits). In a selected basis $\{v\}$ the amplitude errors can be written as 
\beq
A_{\alpha}\ket{v} =\ket{v+\alpha},
\eeq  
where the addition is performed modulo $2$. Analogously, phase errors
are defined as a sequence of $\sigma_z$  transformations performed on
qubits at locations specified by a binary n-tuple $\beta$  and can be 
written as
\beq
P_{\beta}\ket{v} = (-1)^{\beta \cdot v}\ket{v},
\eeq
where the addition in the scalar product $\beta \cdot v$ is also
performed modulo $2$. For example, if $\alpha = \beta =(001010)$ 
and $v=(110111)$, then
\beq
A_{\alpha}\ket{110111}=\ket{111101},\;\;P_{\beta}\ket{110111}=
(-1)\ket{110111}.
\eeq
Amplitude and phase errors are generated by unitary operations and
are, of course, different from errors due to the qubit-channel
entanglement, however, codes which can correct both amplitude and
phase errors can also correct the entanglement induced errors.  To
illustrate the basic idea we start with a simple example of
decoherence induced errors which can be rectified by phase correction
alone. Consider the following scenario:  we want to transmit one qubit
in an unknown quantum state  of the form $c_0\ket{0}+c_1\ket{1}$ and
we know that any single qubit which is transmitted via the channel
can, with a small probability
$p$, undergo a decoherence type entanglement with the channel
\begin{equation}
  (c_0 | 0 \rangle + c_1| 1 \rangle) |a \rangle \longrightarrow
    c_0| 0 \rangle |a_0 \rangle + c_1 | 1 \rangle |a_1\rangle,
\label{simdecoh}
\end{equation}
where states $|a\rangle, |a_0\rangle, |a_1\rangle$ are the states of the
environment/channel  and $|a_0\rangle, |a_1\rangle$ are usually not
orthogonal ($\langle a_0| a_1\rangle\ne 0$). It turns out that with a 
simple encoding and phase error
correcting procedure the probability of error can be reduced to be of
the order $p^2$.  To
achieve this the sender can add two qubits, initially both in state 
$\ket{0}$, to the
original qubit and then perform an encoding unitary transformation
\begin{eqnarray}
\ket{000}&\longrightarrow &\ket{C^0}=\ket{000}+\ket{011}+\ket{101}+\ket{110},\\
\ket{100}&\longrightarrow &\ket{C^1}=\ket{111}+\ket{100}+\ket{010}+\ket{001},
\end{eqnarray}
(here and in the following we omit the normalisation factors) generating state
$c_0\ket{C^0}+c_1\ket{C^1}$. Now, suppose that only the first
transmitted qubit became entangled with the channel; the code-vectors 
$\ket{C^0}$ and $\ket{C^1}$ evolve as
\begin{eqnarray}
\ket{C^0}\ket{a} &\longrightarrow &
(\ket{000}+\ket{011})\ket{a_0}+(\ket{101}+\ket{110})\ket{a_1}\\
\ket{C^1}\ket{a} &\longrightarrow &
(\ket{111}+\ket{100})\ket{a_1}+(\ket{010}+\ket{001})\ket{a_0}
\end{eqnarray}
The receiver applies two projection operators to the received triple
of qubits. Projector $L_1$ projects on the subspace spanned by
$\{\ket{C^0}, \ket{C^1}, P_{100}\ket{C^0}, \\
P_{100}\ket{C^1}\}$  and
$L_2$ on the subspace spanned by
$\{\ket{C^0}, \ket{C^1}, P_{010}\ket{C^0}, \\P_{010}\ket{C^1}\}$. If a
state vector is projected on a specified subspace we say that the
result of the projection is 1 and if the vector is projected on an 
orthogonal subspace we call the result 0. There are four
possible results of the two subsequent projections $L_1$ and
$L_2$: when the result is 11 the final state is the original state
$c_0\ket{C^0}+c_1\ket{C^1}$; results  01, 10 and 00 correspond to
final states which are related to the original one respectively via $P_{100}$, 
$P_{010}$, and $P_{001}$. Depending on the result of the projections
we apply one of these three phase correcting unitary operations and
restore the state. This way we can
achieve an error-free communication in cases when one qubit has
decohered and as the
result the probability of the successful transmission increases to $1 - (1-p)^3
-3(1-p)^2p\approx 1-p^2$. The reason why in this particular case the
phase error
correction (i.e. projections on subspaces of the form
$P_{\beta}\ket{C^k}$) can rectify
errors due to decoherence is because the decoherence process described by
Eq.(\ref{simdecoh}) is mathematically equivalent to randomizing phase
$\phi$ in $c_0\ket{0}+c_1e^{i\phi}\ket{1}$~\cite{decoherence}.

Let us now consider the most general dissipation in the channel; each qubit can
undergo the following entanglement
\begin{eqnarray}
\ket{0}\ket{a}&\longrightarrow
&\ket{0}\ket{a_{0,0}}+\ket{1}\ket{a_{0,1}}
\label{dissgen1}\\
\ket{1}\ket{a}&\longrightarrow 
&\ket{0}\ket{a_{1,0}}+\ket{1}\ket{a_{1,1}}.
\label{dissgen2}
\end{eqnarray}
The states of the channel/environment that entangle with the
transmitted qubits are, in
general, different for different qubits.

We will show now that in order to disentangle up to $t$ qubits from
the channel we
need only amplitude and phase correction codes.

Codes which correct up to $t$ amplitude errors are constructed by 
selecting $2^l$
mutually orthogonal code-vectors $\ket{C^k}$ ($k=1,2,\ldots 2^l$) from
the $2^n$
dimensional state space of the $n$ qubits such that
\beq
\bra{C^k}A_{\alpha}A_{\alpha'}\ket{C^l} =\delta_{kl}\delta_{\alpha\alpha'},
\label{amp-cond}
\eeq
for any $\alpha$ and $\alpha'$ which satisfy $\mbox{wt} (\alpha), \mbox{wt}
(\alpha')\le t$, where $\mbox{wt}(x)$, the weight of $x$, is the
number of values
different from 0 in the n-tuple $x$. Projections on subspaces
$H_\alpha$  spanned by
vectors $\{A_\alpha\ket{C^k}; k=1,2,\ldots 2^l\}$ identify the error  locations
$\bar\alpha$, and the correcting operation $A_{\bar\alpha}$ can be applied.

Codes which correct up to $t$ phase errors are constructed by selecting $2^l$
mutually orthogonal code-vectors $\ket{C^k}$ ($k=1,2,\ldots 2^l$) from
the $2^n$
dimensional state space of the $n$ qubits such that
\beq
\bra{C^k}P_{\beta}P_{\beta'}\ket{C^l} =
\delta_{kl}\delta_{\beta\beta'},
\label{ph-cond}
\eeq
for any $\beta$ and $\beta'$ which satisfy $\mbox{wt} (\beta),
\mbox{wt} (\beta')\le t$. Projections on subspaces $H_\beta$  spanned
by vectors $\{P_\beta\ket{C^k}; k=1,2,\ldots 2^l\}$ identify the error
locations $\bar\beta$, and the correcting operation $P_{\bar\beta}$
can be applied. 

In order to correct the entanglement induced errors we will require that the
code-vectors $\ket{C^k}$ are carefully selected to satisfy  the
following condition
\begin{equation}
\bra{C^k}P_{\beta}A_\alpha A_{\alpha'}P_{\beta'}\ket{C^l} =
\delta_{kl}\delta_{\alpha\alpha'}\delta_{\beta\beta'},
\label{tot-cond}
\end{equation}
for all $\alpha$  and $\beta$ such that 
$\mbox{wt}({\mbox{supp}}[\alpha]\cup{\mbox
{supp}}[\beta])\leq t$ (${\mbox{supp}}[x]$ denotes the set of
locations where the n-tuple
$x$ is different from zero).  Both conditions (\ref{amp-cond}) and 
(\ref{ph-cond}) are
particular cases of (\ref{tot-cond}).  The encoding unitary
transformation 
maps the basis
states of the original $2^l$-dimensional Hilbert space into  
$2^l$ states $\{\ket{C^k}\}$
in the enlarged $2^n$-dimensional Hilbert space. To see how the two codes
can disentangle up to $t$ qubits from the channel consider a
particular case when $t=2$
(cases $t>2$ can be proved by a  simple extension of the argument
presented below).

Let us denote by $\ket{(00)}$ a subset (or a superposition) of the
basis states in which
the two qubits affected by the dissipation process described by
Eqs.(\ref{dissgen1})-(\ref{dissgen2}) are initially both in state
$\ket{0}$, and
analogously for $\ket{(01)}$, $\ket{(10)}$ and
$\ket{(11)}$. For simplicity, let us now restrict our attention to one
of the code-vectors
$\{\ket{C^k}\}$, it can be written  as
\beq
\ket{C^k}=\ket{(00)}_1^k+\ket{(01)}_2^k+\ket{(10)}_3^k+\ket{(11)}_4^k.
\label{vect-ck}
\eeq
After the dissipation the state  $\ket{C^k}\ket{a}$ has the form
\begin{eqnarray}
\ket{(00)}_1^k\ket{a_{00,00}}&+&
\ket{(01)}_2^k\ket{a_{01,01}}+
\ket{(10)}_3^k\ket{a_{10,10}}+
\ket{(11)}_4^k\ket{a_{11,11}}+
\nonumber\\
\ket{(01)}_1^k\ket{a_{00,01}}&+&
\ket{(00)}_2^k\ket{a_{01,00}}+
\ket{(11)}_3^k\ket{a_{10,11}}+
\ket{(10)}_4^k\ket{a_{11,10}}+
\nonumber\\
\ket{(10)}_1^k\ket{a_{00,10}}&+&
\ket{(11)}_2^k\ket{a_{01,11}}+
\ket{(00)}_3^k\ket{a_{10,00}}+
\ket{(01)}_4^k\ket{a_{11,01}}+
\nonumber\\
\ket{(11)}_1^k\ket{a_{00,11}}&+&
\ket{(10)}_2^k\ket{a_{01,10}}+
\ket{(01)}_3^k\ket{a_{10,01}}+
\ket{(00)}_4^k\ket{a_{11,00}}.
\label{2errori}
\end{eqnarray}
By expressing each component of (\ref{vect-ck}) as a linear
combination of phase projectors acting on $\ket{C^k}$:
\begin{eqnarray}
\ket{(00)}^k_1&=&\left(1+P_{01}+P_{10}+P_{11}\right)\ket{C^k}\\
\ket{(01)}^k_2&=&\left(1-P_{01}+P_{10}-P_{11}\right)\ket{C^k}\\
\ket{(10)}^k_3&=&\left(1+P_{01}-P_{10}-P_{11}\right)\ket{C^k}\\
\ket{(11)}^k_4&=&\left(1-P_{01}-P_{10}+P_{11}\right)\ket{C^k}
\label{proj}
\end{eqnarray}
(in a more general case this expression can be derived directly from
the Hadamard transformation), we can write the decohered state 
(\ref{2errori}) as
\begin{eqnarray}
\sum_{\alpha\beta} A_\alpha P_\beta \ket{C^k}\ket{R_{\alpha\beta}},
\label{sum-pr-ck}
\end{eqnarray}
where $\mbox{wt}({\mbox{supp}}[\alpha]\cup{\mbox{supp}}[\beta])\leq 2$ and
$\ket{R_{\alpha\beta}}$ is the state of the channel/environment which 
depends on
$\alpha$ and $\beta$ but, nota bene, not on $k$. More precisely, 
$\ket{R_{\alpha\beta}}$
can be written as 
\begin{eqnarray}
\ket{R_{\alpha\beta}}=\sum_{\gamma}
(-1)^{\gamma\cdot\beta}\ket{a_{\gamma,\gamma+\alpha}}
\label{r-ab}
\end{eqnarray}
where $\gamma$ can take the binary values $00, 01, 10, 11$. An arbitrary 
encoded state i.e. a superposition of code-vectors $\ket{C^k}$ of the form
\begin{eqnarray}
\ket{\psi}=\sum_{k=1}^{2^l} c_k\ket{C^k},
\label{gen-enc}
\end{eqnarray}
evolves under dissipation from the state $\ket{\psi}\ket{a}$ to
\begin{eqnarray}
\sum_{\alpha\beta}A_\alpha P_\beta\sum_k c_k \ket{C^k}\ket{R_{\alpha\beta}}.
\label{sum-gen}
\end{eqnarray}
Now projections on orthogonal subspaces $H_{\alpha\beta}$ spanned by
$\{A_\alpha P_\beta\ket{C^k}, k=1,2^l\}$ are performed. 
The results of the projections
identify the error locations $\bar\alpha$ and $\bar\beta$ and the 
appropriate `state
restoring ' transformation $P_{\bar\beta} A_{\bar\alpha}$ is applied. 
We can see from Eq.
(\ref{sum-gen}) that the state after corrections is of the form $\sum_k
c_k\ket{C^k}\ket{R}$, i.e. the $n$ qubits system is completely
disentangled from the
channel/environment. The generalisation to the $t>2$ case is straightforward. 

Thus we have shown that by a suitable choice of the encoding vectors
$\ket{C^k}$, which satisfy condition (\ref{tot-cond}),  and with
amplitude and phase
corrections we can increase the probability of an error-free
communication in a noisy
quantum channel. Let us mention in passing that searching for error locations
$\bar\alpha$ and $\bar\beta$ does not have to involve projections on
$H_{\alpha\beta}$ for all allowed values $\alpha$ and $\beta$. 
This search can be made
efficient by starting with projections on subspaces which are unions of several
$H_{\alpha\beta}$ and by subsequent divisions and projections on
smaller subspaces.

Quantum encoding requires $n-l$ auxilary qubits as an input to the
encoder. We will now
establish bounds on $n$, i.e. number of qubits needed to encode any state of
$l$ qubits. According to what we have shown above, up to $t$ 
entanglement-induced 
errors can be corrected if we can combine two distinct  procedures
which can correct up
to $t$ amplitude and phase errors. 
Amplitude and phase errors correspond respectively to
operations $\sigma_x$ and $\sigma_z$ performed on selected qubits; 
the two operations
performed on the same qubit can be viewed as the third type of error 
corresponding to
operation $\sigma_y$. In order to be able to establish the location
and the type of errors
we require that all the $2^l$ code-vectors $\ket{C^k}$ and all the
states that are
obtained by applying up to $t$ amplitude and/or phase transformations
are mutually
orthogonal. The total number of orthogonal states must be smaller than
$2^n$ which is the dimension of the Hilbert space of $n$ qubits.  
Thus if we have $i$ errors of the three
types $\sigma_x$, $\sigma_y$, and $\sigma_z$ in an $n$-qubits state there are
$3^i\left(\begin{array}{c} n
\\ i \end{array}\right)$ different ways in which they can occur and
the argument based on counting orthogonal states reduces to
\begin{eqnarray}
2^l\sum_{i=0}^t 3^i\left(\begin{array}{c} n \\ i
\end{array}\right)\leq 2^n.\label{hamming}
\end{eqnarray}
Eq. (\ref{hamming}) is the quantum version of the Hamming bound for classical
error-correcting codes~\cite{macw}; given $l$ and $t$ it provides a
lower bound on $n$. 
The quantum version of the classical Gilbert-Varshamov bound~\cite{macw} can be
also obtained:
\begin{eqnarray}
2^l\sum_{i=0}^{2t} 3^i\left(\begin{array}{c} n \\ i
\end{array}\right)\geq 2^n.\label{gvbound}
\end{eqnarray}
This expression can be proved from the observation that in the $2^n$ 
dimensional Hilbert
space with a maximum number of code-vectors $\ket{C^k}$ any vector which is
orthogonal to $\ket{C^k}$ (for any $k$) can be reached by applying up
to $2t$ error
operations of $\sigma_x$, $\sigma_y$, and $\sigma_z$ type to any of the
$2^l$ code-vectors. Clearly all vectors which cannot be reached in the
$2t$ operations
can be added to the code-vectors $\ket{C^k}$ as all the vectors into
which they can be
transformed by applying up to $t$ amplitude and/or phase
transformations are orthogonal to all the others. 
This situation cannot happen because we have assumed that the number
of code-vectors is maximal. Thus the number of orthogonal vectors that
can be obtained
by performing up to $2t$ transformations on the code-vectors must be
at least equal to
the dimension of the encoding Hilbert space.

It follows from Eq.(\ref{hamming}) that protecting one qubit against 
one error (
$l=1$, $t=1$) requires at least $5$ qubits and, according to Eq. 
(\ref{gvbound}), this can be achieved with less than $10$ qubits.  
Indeed, explicit constructions of quantum codes
for $n=9$, $n=7$ and $n=5$ are known~\cite{shor,steane,smolin}.

The asymptotic form of the quantum Hamming bound (\ref{hamming}) in
the limit of large $n$ is given by 
\begin{eqnarray}
\frac{l}{n}\leq 1-\frac{t}{n}\log_2 3 -H(\frac{t}{n}),\label{hamasym}
\end{eqnarray} 
The corresponding asymptotic form for the quantum Gilbert-Varshamov bound
(\ref{gvbound}) is
\begin{equation}
\frac{l}{n}\geq 1-\frac{2t}{n}\log_2 3 -H(\frac{2t}{n}),\label{gvasym}
\end{equation}
where $H$ is the entropy function $H(x)=-x \log_2 x-(1-x)\log_2(1-x)$. 

Our general requirements for quantum error correcting codes 
(Eq. (\ref{tot-cond})) apply to a variety of codes including  
quantum codes based on classical error
correcting schemes  (c.f. constructions proposed by Calderbank and 
Shor~\cite{calshor},
and by Steane~\cite{steane,note}). 
Like in the classical case there is probably no systematic
way to construct good quantum error correcting codes but we hope that criterium
(\ref{tot-cond}) will make future heuristic approaches easier.

Although we have presented the unitary encodings and the decoding 
projections in a
fairly abstract way they can be implemented in practice as a sequence
of quantum
controlled-NOT logic gates~\cite{cnot}. For experimental purposes
gates that operate
directly on carriers of information, such as recently proposed 
implementation of the
controlled-NOT operating directly on polarised photons~\cite{Kimble}, 
seem to be very
well suited for quantum communication. Other possible applications of 
quantum error
correction may involve improving some high precision measurements e.g.
frequency
standards based on trapped ions. Properly encoded quantum states of
ions will be more
robust to dephasing mechanisms such as, for example, collisions with
the buffer gas and
may have much longer lifetime.  Finally let us also point out that the
encoding described
in this paper applies both to pure and mixed states.
In particular it can be used in
distribution of entangled particles because it allows to encode
(and therefore protect
against errors) each particle separately without destroying the
entanglement. 
It may also lead to better quantum cryptographic protocols~\cite{qc}.


\begin{thebibliography}{99}
\bibitem{note} For example, the code-vectors can be constructed from
  selected codewords $\{v_i\}$ as $\ket{C^k}=\sum_i\ket{v_i^k}$.
  Requirement (\protect{\ref{amp-cond}}) implies that  for any
  $\alpha$ and 
$\alpha'$ (both of weight less than $t$) $v^k_i+\alpha +\alpha'\ne
v^l_j$. This means that the selected codewords must be
separated at least by the Hamming distance  $2t+1$. Requirement  
(\ref{ph-cond})
implies that $\sum_i (-1)^{v^k_i \cdot
  (\beta+\beta')}=\delta_{\beta\beta'}$ for any $k$, $\beta$ and
$\beta'$ ($\mbox{wt}(\beta),\mbox{wt}(\beta')\le t$). 
If for a given $k$ the codewords
$\{v_i^k\}$ form a linear code $\cal C$ then this condition is
satisfied when the dual
code $\cal C^\perp$ has minimum distance $2t+1$.
\end{thebibliography}
\end{document}